# Decision Sort and its Parallel Formulation

Udayan Khurana
*Computer Science and Engineering*
*Thapar Institute of Engineering and Technology*
*Patiala, Punjab, India- 147004*
udayankhurana@gmail.com

## Abstract

*In this paper, a sorting technique is presented that takes as input a data set whose primary key domain is known to the sorting algorithm, and works with an time efficiency of O(n+k), where k is the primary key domain. It is shown that the algorithm has applicability over a wide range of data sets. Later, a parallel formulation of the same is proposed and its effectiveness is argued. Though this algorithm is applicable over a wide range of general data sets, it finds special application (much superior to others) in places where sorting information that arrives in parts and in cases where input data is huge in size.*

## 1. Introduction

In this age of information and in the era of distributed computing, one thing that is growing at an exponential rate is information itself. As more and more nodes become a part of the Internet community, data sharing grows as never before. Every millisecond, thousands of new tables are created and older ones updated all over the world. With electronic storage becoming a household affair, new data is created and older one replicated and specialized for individual use. Hence, it becomes a necessity to arrange data in a manner that it is retrieved cost effectively. Sorting records is thus an omnipresent activity. Because of such large amount of data of all types, it is not a bad idea to formulate techniques that which, though applicable to a limited class of data, perform better than a generic technique. In this paper, a new sorting technique is presented and its wide scope of applicability is discussed. The pre requisite for the algorithm to function on the input data set is the knowledge of the range of data set or the primary key (for we sort the data according to the primary key). This range is from here on referred to as the primary key domain (k).The knowledge of the upper and lower bound of the possible values is sufficient to determine the primary key domain. We need an additional binary string, sized k, which we call the Auxiliary Decision String (ADS), the need of which shall be clear as one reads on. The technique is a special one because of its fixed time requirement independent of the actual data, though it varies with the size and range of data. The same should be useful in predicting the actual time before the actual sorting starts. This technique uses no swapping mechanism but only read, write and compare operations. Benefits of the same are also discussed. From here on, this sort shall be referred to as Decision Sort, because of the nature of the algorithm.

Sorting techniques have a long list. From the $O(n^2)$ historical techniques like Bubble, Selection and Insertion Sort[1,2,9], there have been much effective sorts like Radix, Bucket, Shell [4], Merge, Quick[3], Heap Sort[5]. Also, there have been many variants of all these sorts that specialize them to give enhanced results. E.g. [6,9] American Flag Sort is a kind of Radix Sort. However, the once named earlier parent almost all types of basic sorting techniques used. Most of those implemented today are O(nlogn) ones. The various flavours of all of those may be found at [9].

## 2. Basic Thought

The basic thought behind this algorithm may be stated as follows. The algorithm is composed of two basic components. We assume to know the upper and lower extrema and hence the possible domain of the key. We have with us a binary string of size equal to the key domain, initialized to zero. The first component of the algorithm would include a scan of the whole list of the key set, and marking those indexes of the binary string high which correspond to all the values in the key set. This gives us the precious information regarding the presence or absence of an element in the key set. The second component of the algorithm scans through the entire binary string and for all entries marked one, writes in the array, the corresponding element. This involves iterations as many as the size of the range. Here, the algorithm presented is one taking input values as numerical data. It can be easily extended in case a categorical attribute is the sorting key. Also, since the primary key is unique for all records in a set, we assume that each instance of a value occurs only once. If the requirements in a practical problem do not obey the same, this algorithm can be easily extended by taking the binary string alphabets as general integral ones. $E_l$ and $E_h$ denote the lower and upper limit of input data set, L.

## 3. Algorithm

Algorithm DecisionSort(L,$E_l$,$E_h$,k)

ArraytobeSorted : L
Lower Limit : $E_l$
Upper Limit : $E_h$
Size : k
Binary String ADS(k)
i=0
while( not end of L)
    ADS[L[i]-$E_l$ ]=TRUE
    i++
end while

i=0, j=0
while( not end of ADS)
    if(ADS[i]==TRUE)
        L[j]=i+$E_l$
        j++
    end if
    i++
end while
end DecisionSort

NOTE: The steps including L[I]-$E_l$ and I+$E_l$ are nothing but mapping of indexes from L to ADS and vice versa respectively.

## 4. Algorithm Analysis

### 4.1 Complexity

The simple and easy algorithm for decision Sort is even simpler to analyze.
It consists of two while loops that make up the entire set of steps. So, let us analyze the two one after one.

The first One
It can be easily seen that the loop iterates fixed number of times equal to the size of the array to be sorted. It does not vary and the number of iterations=n

The second One
We see that the second loop also iterates a fixed number of times, i.e.=k. The loop consists of a decision statement which is true n number of times (fixed), since it comes true only when the particular element in ADS is High or True and it is known that ADS consists of n number of TRUE bits. Hence, the total cost = n+k.

Combining the steps as discussed above, we have the total number of iterations as
    $I_{total}$ = n+k, fixed     (1)
If we take into care the total number of steps that are executed, we get
    $S_{total}$ = 2n+k, fixed     (2)

Hence, we may want to say that the time complexity of the algorithm is of the order of n+k, in the best worst and average case.
    Complexity = O(n+k)     (3)

### 4.2 No Swapping

An important point to note here that the algorithm does not involve swapping of any elements. Only read, write or compare operations are required. A swap is generally costlier to implement than a read/write or compare step. This suggests that the cost of proportionality (and hence the proportionality constants) are low.

### 4.3 Optimality of Decision Sort

The order of time complexity of decision sort is variable with the size parameter k, since the complexity term contains the sum of n and k. Let us look at the various possible cases. One must bear in mind that k is always greater than equal to n.
    Case:    k $\alpha$ n
    In this case, the complexity is O(n).

Case: $k \propto n^a$, $a > 1$
In this case, the complexity is $O(n^a)$
Case: $k \propto n.\log(n)$
In this case, the complexity is $O(n \log n)$
(4)

Out of the three, first and third case are acceptable everywhere, whereas the second one is good as long as the value of a is less than some desired constant. It is known by experience that for the values of a < 2 or a < 1.7 should be contenders for an optimal sorting algorithm

It may be interesting to note that the complexity always has a very low constant of proportionality, so heuristics should suggest even an algorithm with lesser order should do worse than this in many real size problems.

Let us look at some real size problems and try to make an assessment of the applicability of Decision Sort. The value in the bracket shows the proportion of range in terms of the data input size. The inverse of the same would suggest the probability of finding an element of L in the range k.

| Problem Size(n) | Range(k) | Exponent (a) |
|---|---|---|
| 100 | 400 (4 times) | 1.3 |
| 1000 | 10000 (10 times) | 1.3 |
| 100000 | 10000000 (100 times) | 1.4 |
| 100000000 | 10000000000 (100 times) | 1.25 |
| $10^{16}$ | $10^{22}$ (1 million times!) | 1.37 |

Table 1. Range and Problem Size

Taking into consideration instances in the above table, in the first case, we can interpret that if the data size be100 and the range of elements be four times, i.e. the size of k=400, (a fair approximation) we will have an optimal run by the Decision Algorithm. As the size increases, so does the permissible range for some fixed a. For greater values of n, the algorithm becomes more and more suitable in practice.

**4.4 Range-Probability Trade-Off**

This subsection gives an account of the trade-off between permissible range and probability for a fixed value of a for data set that has the relation ship as:
$k \propto n^a$.
This is actually an unusual trade-off, as it limits probability (remember, lesser the probability, the better) with lesser data sizes, for a fixed a. Though, this is not relevant, as in the real time problems, one would have the provision of both n and k and the calculation of exponent a shall decide the optimality(maybe, hence the use) of this algorithm.

Since, $k = c.n^a$ (for only polynomial case)
$\Rightarrow k/n = n^{a-1}$
Taking Probability, $P = n/k$
We have, $P . n^{a-1}$ = constant, (5)
which represents a hyperbolic function such as one given in the Figure 1.

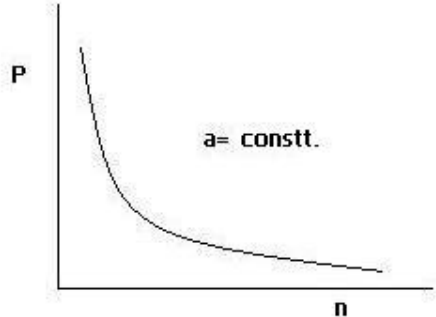

Figure1. P v/s n

## 5. Illustration, Comparison and Application

### 5.1 An Example

Let us take an example to illustrate the Decision Sort algorithm and compare it to two historic algorithms. While comparison, we take the necessary assumption that a swap step is three times as costly as a read write or compare step. A swap may be shown to consist of at least three write operations.

Let there be an unsorted data with key set,
L = {4,2,7,9,1,13,15}, n= 7
Therefore, $E_l = 1$, $E_h = 15$.

And, size of the Auxiliary Decision String (ADS),
$k = E_h - E_l + 1 = 15 - 1 + 1 = 15$

Stage I
Following the algorithm, the state of ADS would be:
```
            Element         ADS
Iteration 1:    4       000100000000000
Iteration 2:    2       010100000000000
…….            .         …………………..
Iteration n:   15       110100101000101
```

Number of writing steps=n
Number of elementary steps in stage I = number of steps in (1) n=7   (6)

Stage II

Following the algorithm, the second stage, would follow k=15 comparisons amongst which n=7 are successful and the remaining n-k=8 unsuccessful. The following show, a few iterations:

     State Of List L
Iteration 1:  {1}
Iteration 2:  {1,2}
Iteration 3:  {1,2}
Iteration 4:  {1,2,4}
…..
Iteration k:  {1,2,4,7,9,13,15}  *Sorted List*

Number of comparisons= k = 15   (7)

Number of writing steps = n = 7   (8)

Number of elementary steps in stage II
  = Sum of (7) & (8)
  = k+n=15  (9)

Hence, total number of elementary steps in the whole sorting procedure = Sum of (6) & (9) = 15 + 7 =22.
           (10)

Table 2 compares results with some other sorting techniques in terms of the number of elementary steps that are required in practice to execute. A swapping step is weighed three times a read/write/comparison step and the "total number" indicates relative weights.

| Sort | Comparison/ Writing | Swaps | Total Number (3S+C/W) |
|---|---|---|---|
| Bubble | 15 | 15 | 60 |
| Quick | 15 | 5 | 30 |
| Decision | 22 | - | 22 |

Table 2. Comparison

Among the three, Decision Sort would thus take the least computational time.

## 5.2 Applications

### 5.2.1 Some cases of application

 To complement the algorithm, few instances of application are suggested as follows. A case where sorting is to be done for a data set of students as per the merit in one or more subjects. Here the possible range of marks is known, lets say 100 or 200, and to sort a set of 70 students according to their marks or grades.

### 5.2.2 An Interesting Application

Let us consider a situation where a machine is supposed to sort values and transport it, but the data come in parts and not at once. A typical sorting algorithm would wait for the complete data set to arrive or sort data as it comes, maybe use some insertion sort kind of an algorithm. In the first case, the processor may be idle for too large amount of timings and the sorting submission would also be late. The second is a better choice, but Decision sort would be much more effective (time saving) in such a case.

## 6. Parallel Formulation

### 6.1 Basic Idea

It seems quite interesting that the Decision Sort can be easily implemented on a parallel system. Giving a naïve thought can suggest use of p parallel processors (all with binary strings of full size) assigned n/p blocks of input data each. Every processor takes n/p steps to complete the stage one job. In the second stage, each a broadcast is done such that each processor receives a copy of the binary string and XoRs it with the latest copy it has, while it waits for strings from other processors. Finally, each processor processes k/p sized data set for the second stage. And then, may be a single node accumulation or a broadcast would do to collect the sorted data at one node or all of them respectively. No additional sorting is required, since the sorted data with each processor is in the order of their labels. i.e. all the values with processor I are smaller than those with processor j for all i <j.

### 6.2 Assessment

Considering a logp dimensioned hypercube formulation, we have:
The total number of steps,
$T = n/p + \log p + k/p + \log p = n/p + 2\log p + k/p$. (11)

Hence, the processor-time product,
$R = p*(n/p + 2\log p + k/p) = n + p\log p + k$.  (12)

For this to be a good implementation, we should have the following constraint apart from the ones for the serial version;
  $2p\log p \; \alpha \; (k)^a$  where a <1
(13)

Equation (13) suggests that 2plogp should have a lesser order than k.

So, for 2plogp $\alpha$ (k)[a], this technique should be parallelizable, which seems a case that shall occur quite frequently.

Its speedup can be calculated as:
$$S = (n + k)/ (n/p + 2\log p + k/p)$$
$$= p (n+k)/(n+k+2p\log p) \quad (14)$$

So, this approaches p, for relatively insignificant values of plogp.
Also, efficiency, $E = 1/(1 + 2p\log p/(n+k))$ (15)

Let's take the case for some values of n, k and p in Table 3.

| N | K | p | S | E |
|---|---|---|---|---|
| 100 | 400 | 8 | 7.3 | 0.92 |
| 1000 | 5000 | 8 | 7.94 | 0.99 |
| 100000 | 1000000 | 16 | 15.998 | 0.9999 |

Table 3. Speedup, Efficiency v/s n,k,p

## 7. Conclusion and Scope for Further Research

In this paper, we have presented a sorting technique that works in quite a strange manner, i.e. there are no comparisons between two elements of the input list. But every possible element is checked for actual presence and marked as absent or present. The list of presentees is thus produced as the sorted list. The technique is quite cheap for a large domain of input data size and range and shows fixed behaviour (time taken) for a specific input size and range, irrespective of the data and its ordering. Later, it was suggested that its parallel formulation would be quite effective and at place where the input data is received in parts and not altogether. At such a case, the other typical sorting techniques would effectively start processing only when the complete data set has arrived.

Scope for further research may be seen under the following categories:

→ In cases where the exact value of k is not known, but can be decided by historical data, considering all the statistics. Even, approximation methods (determining k) can be found out where data is important only for aggregate information, so loss of some data might be acceptable. Many k-Anonymity [7] and other data masking methods may suggest that.

→ Actual parallel implementation using various parallel computing models and proposition of better parallel formulation.

→ Finding out more and more fields of application where this technique outperforms others.